\documentclass[twocolumn,showpacs,preprintnumbers,amsmath,amssymb]{revtex4-1}

\usepackage{afterpage}
\usepackage[dvipdfmx]{graphicx}
\usepackage{color,bm}

\begin{document}

\title{Hemisphere-averaged Hellings-Downs curve between pulsar pairs for a gravitational wave source}
\author{Tatsuya Sasaki}
\email{sasaki@tap.st.hirosaki-u.ac.jp}
\author{Kohei Yamauchi}
\email{yamauchi@tap.st.hirosaki-u.ac.jp}
\author{Shun Yamamoto}
\email{yamamoto@tap.st.hirosaki-u.ac.jp}
\author{Hideki Asada} 
\email{asada@hirosaki-u.ac.jp}
\affiliation{
Graduate School of Science and Technology, 
Hirosaki University,
Hirosaki 036-8561, Japan} 

\date{\today}

\begin{abstract}
The Hellings-Downs (HD) curve plays a crucial role 
in search for nano-hertz gravitational waves (GWs) with pulsar timing arrays. 
We discuss the angular pattern of correlations for pulsar pairs within a celestial hemisphere. 
The hemisphere-averaged correlation curve depends upon the sky location of 
a GW compact source like a binary of supermassive black holes. 
If a single source is dominant, 
the variation in the hemisphere-averaged angular correlation is greatest 
when the hemisphere has its North Pole at the sky location of the GW source. 
Possible GW amplitude and source distance relevant to the current PTAs 
by using the hemisphere-averaged correlation 
are also studied. 
\end{abstract}

\maketitle

\section{Introduction}
Decades ago, 
the idea of using radio pulse timing to search for gravitational waves (GWs) 
was proposed 
\cite{Estabrook, Sazhin, Detweiler}.
Notably, Hellings and Downs (HD) 
pointed out that the angular correlation pattern for pulsar pairs 
averaged over the sky 
can provide an evidence of GWs 
\cite{Hellings}, 
since the correlation curve is a consequence of 
the quadrupole nature of GWs 
\cite{CreightonBook, MaggioreBook, Anholm2009, Jenet}. 
Along this direction, 
several teams of pulsar timing arrays (PTAs) have recently 
reported an evidence of nano-hertz GWs 
\cite{Agazie2023, Antoniadis2023, Reardon2023, Xu2023}. 
According to their results, 
superpositions of supermassive black hole binaries (SMBHBs)
are among possible GW sources, 
though the origin of the significant correlation has not been identified yet. 
See e.g. Reference 
\cite{Romano} 
for a review on detection methods of stochastic GW backgrounds. 

For an isotropic stochastic background of GWs 
composed of the plus and cross polarization modes 
in general relativity, 
the expected correlated response of pulsar pairs follows the HD curve, 
where the averaging over the whole sky ($4\pi$) is taken. 
The whole-sky average makes the HD curve 
insensitive to any particular direction of the sky. 
This is the price of extracting tiny GW signals from large fluctuations of 
pulse signals from individual pulsars. 
Therefore, it is not surprising that 
an isolated SMBH binary produces an identical cross-correlation pattern 
as an isotropic stochastic background 
\cite{Cornish}.

What happens in the pulsar correlation 
if the averaging domain is changed from the whole sky? 
Any deviation from the the whole-sky average 
breaks the isotropy. 
It may thus allow to use modified cross-correlation curves 
for the sky localization of a GW source. 

The main purpose of this paper 
is to examine if 
the angular correlation pattern for pulsar pairs within a sky hemisphere 
has a dependence on a GW compact source. 
We show that, 
if a  single GW source is dominant, 
the variation in a hemisphere-averaged angular correlation curve is greatest  
when the hemisphere has its North Pole at the sky location of the GW source. 

This paper is organized as follows. 
Section II considers pulsar pairs within a sky hemisphere, 
for which the pulsar correlation curve is discussed. 
In section III, 
we discuss the dependence of 
the maximum correlation and the minimum one 
on the North Pole of a hemisphere. 
We shall show that 
the difference between the maximum and minimum correlations  
for a single hemisphere is greatest,  
if and only if the hemisphere has its North Pole at the sky location of the GW source. 
Section VI is devoted to Conclusion. 
Throughout this paper, 
$a$ and $b$ label two pulsars as a pair, 
and 
the center of the coordinates is the solar barycenter, 
safely approximated as the Earth.

\section{Hemisphere-averaged angular-correlation pattern}
\subsection{Full-sky cross-correlation}
Following Reference 
\cite{CreightonBook, MaggioreBook, Anholm2009}, 
let us discuss a correlation of pulsar pairs. 
See e.g. Appendix C of \cite{Anholm2009} 
for detailed calculations of deriving the standard HD curve.

\begin{figure}
\includegraphics[width=8.0cm]{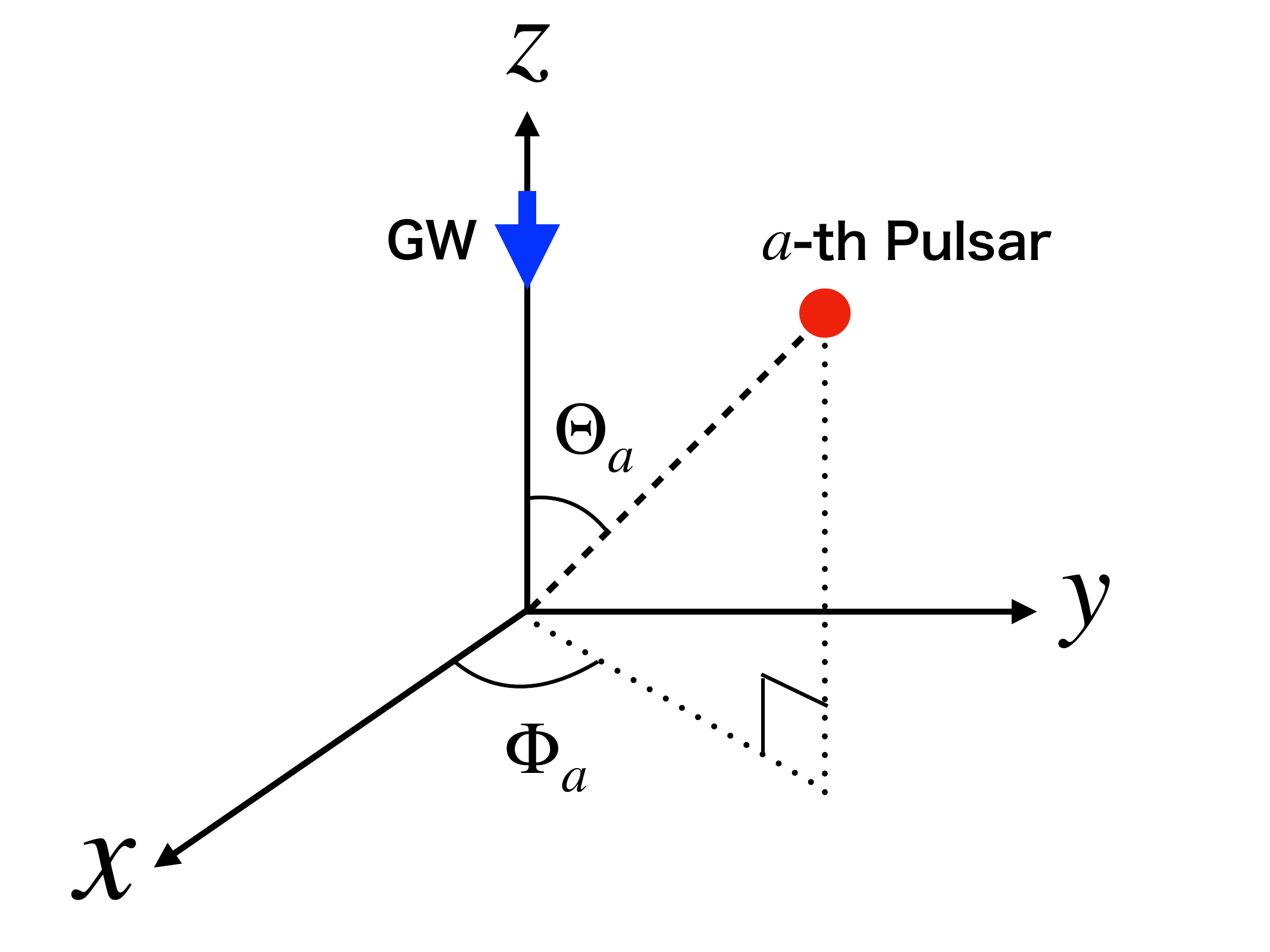}
\caption{
GW-oriented coordinates $(x, y, z)$. 
The GWs propagate downward along the $z$-axis. 
The latitude and longitude of the $a$-th pulsar 
are $\Theta_a$ and $\Phi_a$, respectively. 
}
\label{figure-HD-Config}
\end{figure}

We consider a GW compact source 
which is chosen as the $z$ direction in the coordinates $(x, y, z)$. 
See Figure \ref{figure-HD-Config} for the GW-oriented coordinate system $(x, y, z)$. 
The fractional frequency shift in the timing observation of the $a$-th pulsar 
can be written as 
\cite{Hellings}
\begin{align}
\frac{\delta \nu_a}{\nu_a} 
= R_a h(t) + n_a(t) , 
\label{redshift}
\end{align}
where 
$h(t)$ is the GW signal, 
$n_a(t)$ is the noise, 
and the response 
$R_a =\cos2\Phi_a (1 + \cos\Theta_a)$ 
for $\Theta_a \in [0, \pi)$ and $\Phi_a \in [0, 2\pi)$.

By taking the average over the sky, 
the cross-correlation of pulsar pairs is 
\cite{Hellings, CreightonBook, MaggioreBook, Anholm2009}
\begin{align}
\Gamma_{\rm{S}}
\equiv 
\frac{1}{4\pi} 
\int_S  R_a R_b d\Omega , 
\label{HD}
\end{align}
where $S$ is the unit sphere and the noise terms are assumed to be uncorrelated. 
So far, 
Eq. (\ref{HD}) has been considered 
mostly for isotropic stochastic GW background 
\cite{CreightonBook, MaggioreBook, Anholm2009, Jenet, Agazie2023, Antoniadis2023, Reardon2023, Xu2023}.

\subsection{Hemisphere cross-correlation}
\begin{figure}
\includegraphics[width=8.0cm]{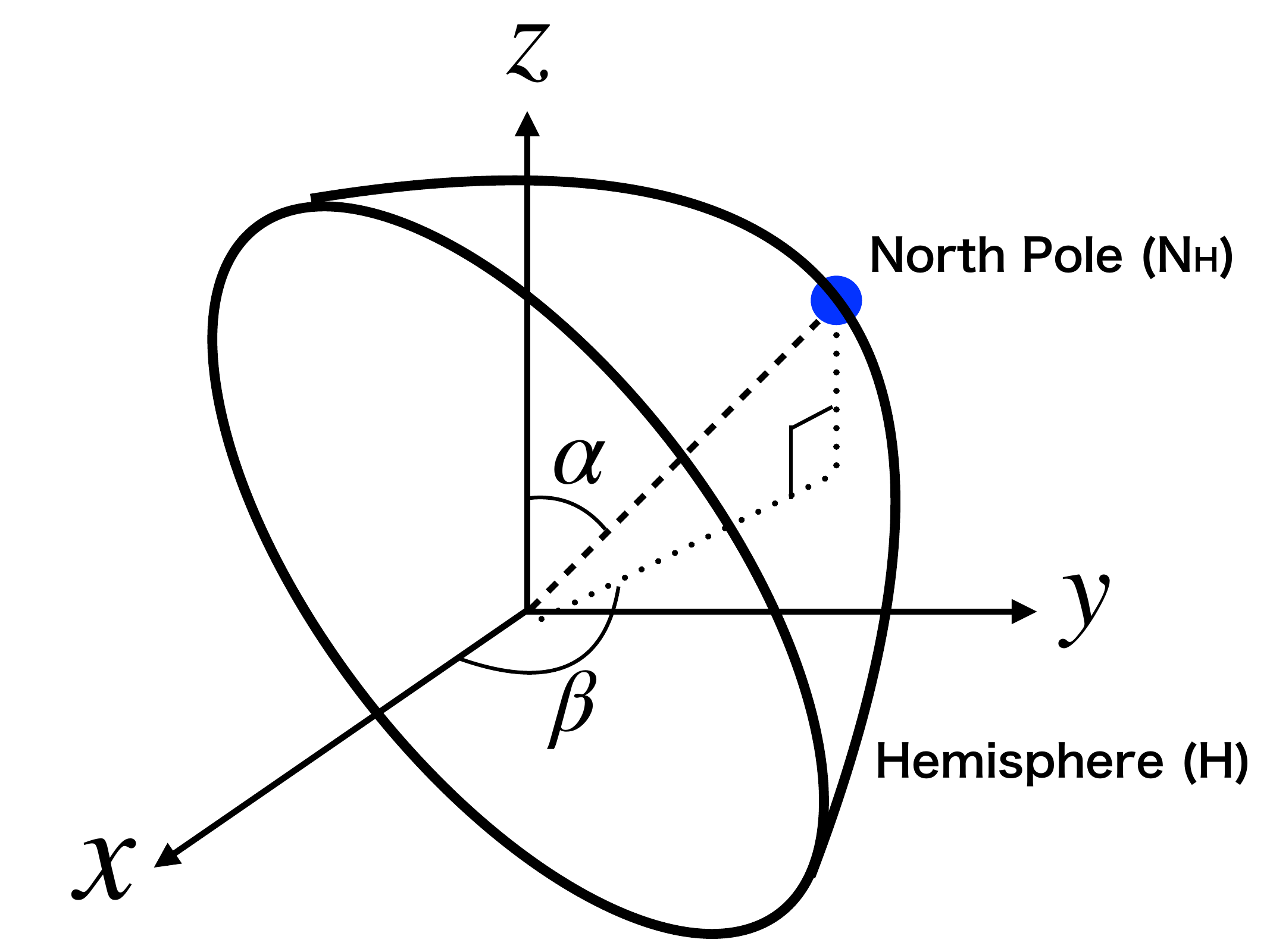}
\caption{
Hemisphere in the coordinates $(x, y, z)$. 
The latitude and longitude of the North Pole 
are $\alpha$ and $\beta$, respectively. 
}
\label{figure-Hemisphere}
\end{figure}

Next, we consider a sky hemisphere $H$ 
in Figure \ref{figure-Hemisphere}. 
The North Pole $\rm{N}_{\rm{H}}$ is 
denoted as $\bm{N}_{\rm{H}} \equiv (\sin\alpha\cos\beta, \sin\alpha\sin\beta, \cos\alpha)$ 
and $\alpha$ is the inclination angle of the hemisphere from the GW source direction. 
Averaging over this hemisphere 
may be expressed as 
\begin{align}
\Gamma_{\rm{H}} 
\equiv 
\frac{1}{2\pi} 
\int_H  R_a R_b d\Omega , 
\label{HD-H1}
\end{align}
where a factor in front of the hemisphere integral is $1/2\pi$.

A practical problem is how to perform the integration, 
because 
$\Theta_a$ and $\Phi_a$ for the hemisphere case
are allowed in a non-trivial domain, 
and thereby calculations of 
$\int_{\rm{H}} d\Omega = \int\sin\theta d\theta \int d\phi$ 
are not straightforward, 
because 
$\Theta_a$ and $\Phi_a$ 
do not respect the North Pole of the hemisphere.  
Therefore, the angle coordinates $\theta$ and $\phi$ 
shall be specified below to respect  
a hemisphere that we wish to study.

\begin{figure}
\includegraphics[width=8.0cm]{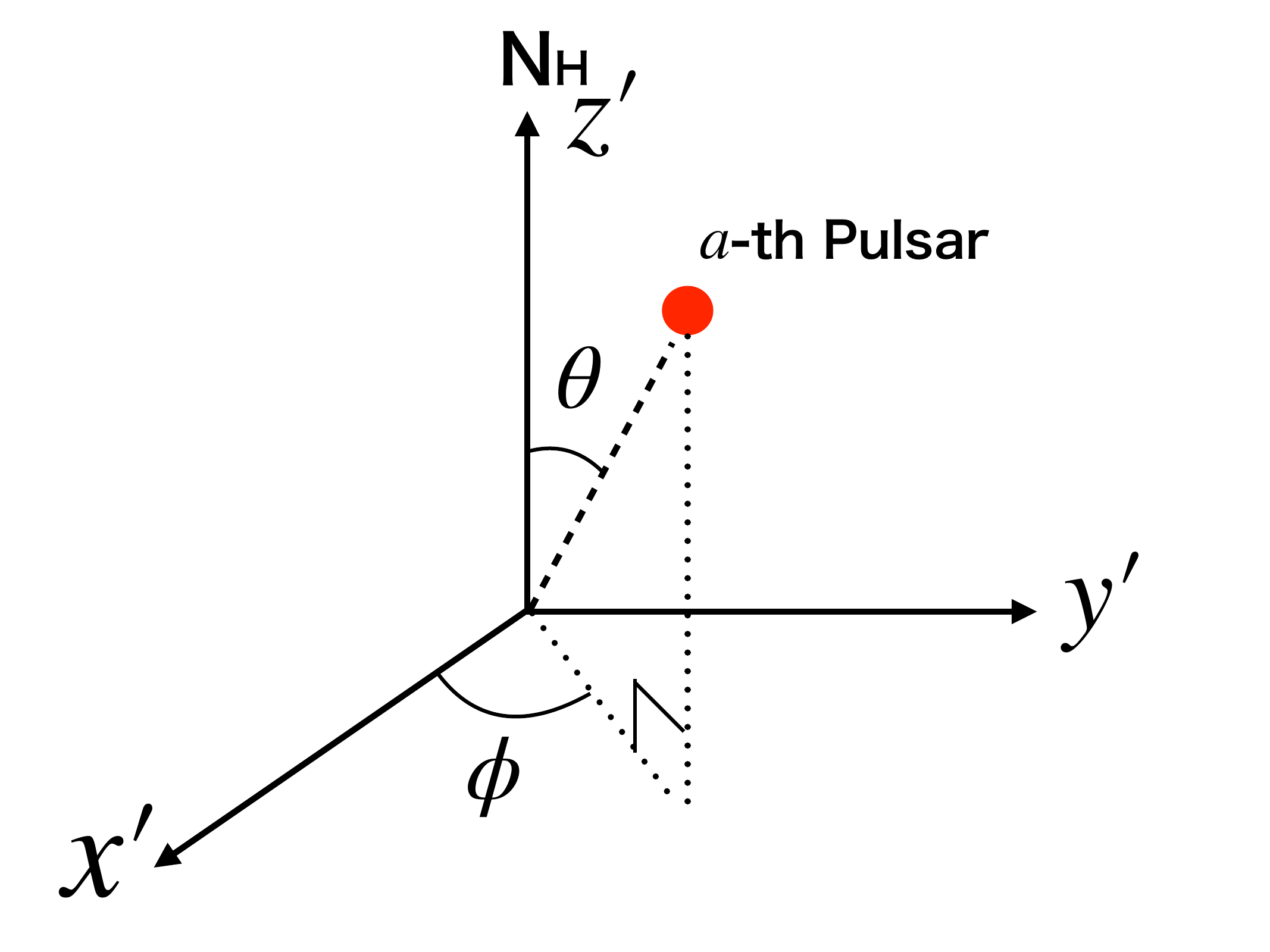}
\caption{
Hemisphere-oriented coordinates $(x', y', z')$. 
The latitude and longitude of the $a$-th pulsar 
are $\theta$ and $\phi$, respectively. 
}
\label{figure-Pulsar-Config}
\end{figure}

For the purpose of practical calculations of Eq. (\ref{HD-H1}), 
therefore, 
it is convenient to introduce another coordinates $(x', y', z')$ 
such that 
the North Pole is chosen as the $z'$-axis.  
See Figure \ref{figure-Pulsar-Config} for the new coordinates 
respecting the North Pole of a hemisphere. 
In the coordinates, 
the unit vector to the $a$-th pulsar can be written as 
$\bm{n}_a = (\sin\theta\cos\phi, \sin\theta\sin\phi, \cos\theta)$, 
for which 
the half sphere as $\theta \in [0, \pi/2)$ and $\phi \in [0, 2\pi)$ 
fully agrees with the hemisphere.

\subsection{Pulsar-oriented coordinates}
\begin{figure}
\includegraphics[width=8.0cm]{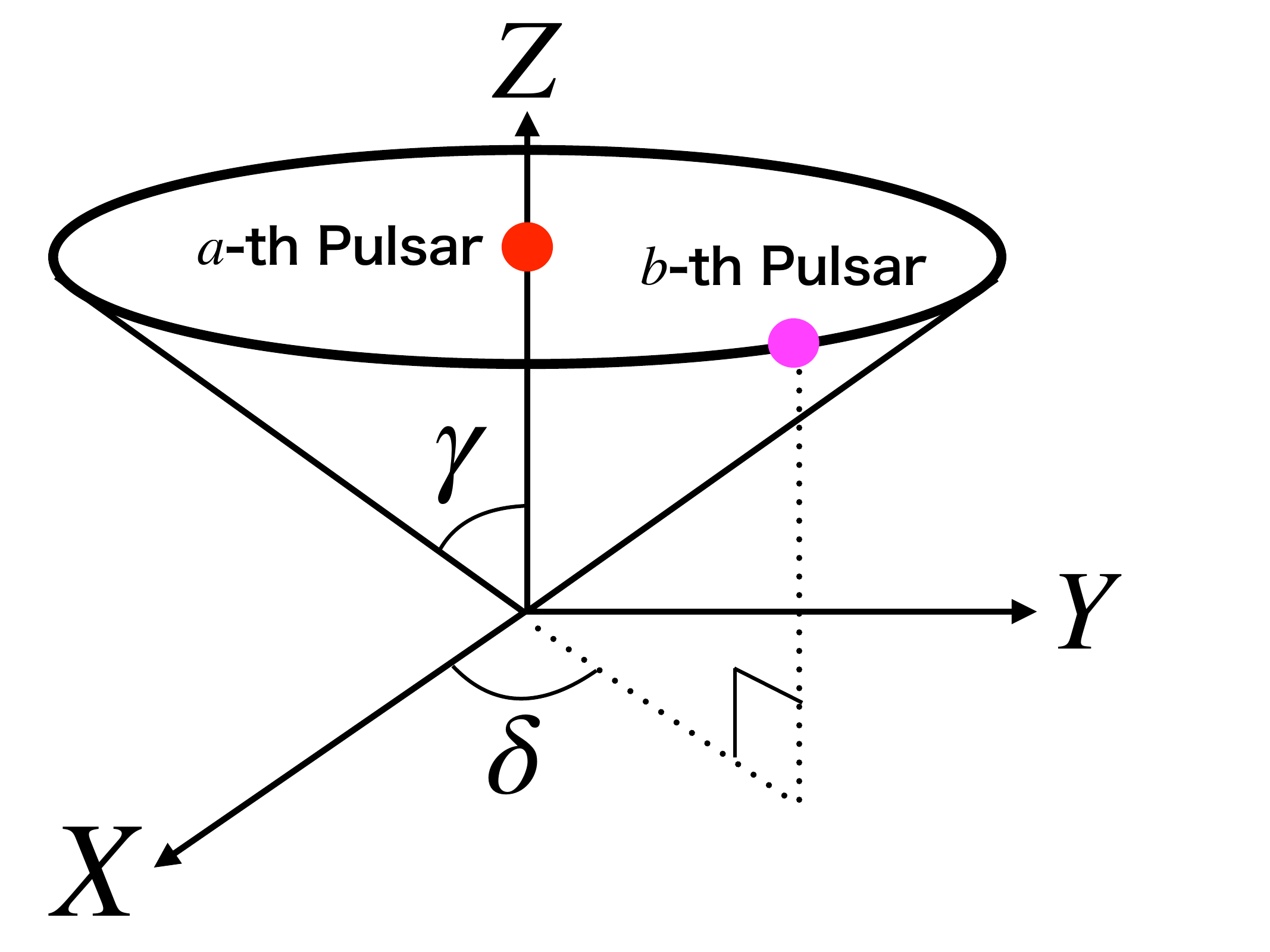}
\caption{
Pulsar-oriented coordinates $(X, Y, Z)$ and 
the separation angle $\gamma$ between 
the $a$-th pulsar and the $b$-th one. 
The $b$-th pulsar is on a cone and its direction is specified by 
two angles $\gamma$ and $\delta$. 
}
\label{figure-Cone-Config}
\end{figure}

The separation angle between the pulsar pair 
is denoted as $\gamma$. 
For clearly describing the configuration of the two pulsars, 
we introduce a third coordinate system $(X, Y, Z)$ 
such that the $a$-th pulsar location is chosen along the $Z$-axis. 
See Figure \ref{figure-Cone-Config} for the coordinates $(X, Y, Z)$. 

The $b$-th pulsar is located on a cone where  
the apex is the origin of the $(X, Y, Z)$ coordinates, 
the apex angle is $\gamma$, 
and the $a$-th pulsar is on the axis. 
In the third coordinate system, 
the $b$-th pulsar direction can be written as 
$\bm{n}_b = (\sin\gamma\cos\delta, \sin\gamma\sin\delta, \cos\gamma)$.

\subsection{Transformations among the GW-oriented, Pulsar-oriented and Hemisphere-oriented coordinates} 
The coordinate transformation from 
$(x', y', z')$ to $(x, y, z)$ 
is a composed rotation as 
\begin{align}
S = 
\begin{pmatrix}
\cos\alpha\cos\beta & -\sin\beta & \sin\alpha\cos\beta \\
\cos\alpha\sin\beta & \cos\beta & \sin\alpha\sin\beta \\
-\sin\alpha & 0 & \cos\alpha 
\end{pmatrix} ,
\label{S}
\end{align}
and that from $(X, Y, Z)$ to $(x', y', z')$ is 
\begin{align}
T = 
\begin{pmatrix}
\cos\theta\cos\phi & -\sin\phi & \sin\theta\cos\phi \\
\cos\theta\sin\phi & \cos\phi & \sin\theta\sin\phi \\
-\sin\theta & 0 & \cos\theta
\end{pmatrix} .
\label{S}
\end{align}

In the $(x, y, z)$ coordinate representation, 
the $a$-th and $b$-th pulsar directions are 
$\bm{n}_a = S (\sin\theta\cos\phi, \sin\theta\sin\phi, \cos\theta)^{\mbox{T}}$, 
and 
$\bm{n}_b = S T (\sin\gamma\cos\delta, \sin\gamma\sin\delta, \cos\gamma)^{\mbox{T}}$, 
where the superscript T denotes the transposition.

\subsection{Hemisphere cross-correlation II}
The response $R_A (A=a, b)$ for the plus and cross modes 
of GWs coming from the direction $\bm{\Omega}$ 
can be rewritten in the covariant form as 
\cite{Anholm2009, Romano}
\begin{align}
R_A^{+} 
&=
\frac12
\frac{(\bm{n}_A\cdot\hat{\bm{\ell}})^2 - (\bm{n}_A\cdot\hat{\bm{m}})^2}
{1 + \bm{\Omega}\cdot\bm{n}_A} 
\notag\\
 R_A^{\times} 
&=
\frac{(\bm{n}_A\cdot\hat{\bm{\ell}})(\bm{n}_A\cdot\hat{\bm{m}})}
{1 + \bm{\Omega}\cdot\bm{n}_A} ,
\label{R2} 
\end{align}
where 
$\hat{\bm{\ell}} = (0, -1, 0)$ and $\hat{\bm{m}} = (-1, 0, 0)$ 
in the the $(x, y, z)$ coordinates 
are the orthonormal bases on the plane perpendicular to the GW propagation 
and 
$\hat{\bm{\ell}}$ and $\hat{\bm{m}}$ can be used 
in the definition of the plus and cross modes.

Therefore, 
the total correlation by both the plus and cross modes is 
\cite{Anholm2009, Romano}
\begin{align}
\Gamma_{\rm{H}}(\alpha, \beta, \gamma)
= 
\frac{1}{(2\pi)^2} 
\int_0^{2\pi} d\delta 
\int_H  (R_a^{+} R_b^{+} + R_a^{\times} R_b^{\times}) d\Omega ,
\label{HD-H2}
\end{align}
where the $a$-th pulsar is averaged on the hemisphere 
($(1/2\pi)\int d\Omega$), 
and 
the $b$-th pulsar is averaged over the cone 
($(1/2\pi)\int d\delta$). 
The denominator in front of the integrals 
is $2\pi (\rm{hemisphere}) \times 2\pi (\rm{\delta\;integral}) = 4\pi^2$.

After substituting Eq. (\ref{R2}) into Eq. (\ref{HD-H2}), 
the integrand is a function of six angles $\alpha, \beta, \gamma, \delta, \theta, \phi$. 
For the average over the present hemisphere, 
Eq. (\ref{HD-H2}) explicitly becomes 
\begin{align}
&\Gamma_{\rm{H}}(\alpha, \beta, \gamma)
\notag\\
=& 
\frac{1}{4\pi^2} 
\int_0^{2\pi} d\delta 
\int_0^{2\pi} d\phi
\int_0^{\pi/2} \sin\theta d\theta
(R_a^{+} R_b^{+} + R_a^{\times} R_b^{\times}) . 
\label{HD-H3}
\end{align}

In the present formulation, 
we do not need take special care of the integration domain, 
whereas Eq. (\ref{HD-H1}) needs a special care for a hemisphere case 
because the GW propagation axis is not aligned with 
the polar axis of a hemisphere.

In this paper, 
the integration in Eq. (\ref{HD-H3}) 
is done numerically for a hemisphere, 
whereas the correlation integration can be done analytically for the full sky
\cite{Hellings, Anholm2009}.

\begin{figure*}
\includegraphics[width=\textwidth]{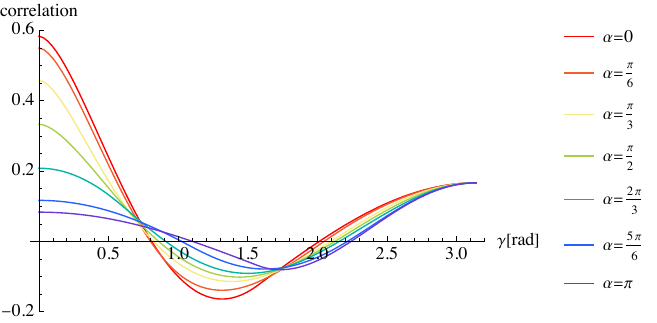}
\caption{
Hemisphere-averaged angular-correlation pattern for pulsar pairs. 
It depends upon an inclination angle $\alpha$, 
which runs from $0^{\circ}$ to $180^{\circ}$ every 30 degrees. 
The separation angle viewed from the Earth is $\gamma$. 
}
\label{figure-curve}
\end{figure*}

\section{GW sky location and the maximum and minimum correlations}
\subsection{Angular correlation pattern for a hemisphere}
Numerical calculations of Eq. (\ref{HD-H3}) give 
angular correlation patterns. 
It follows that the curves depend upon $\alpha$ but not upon $\beta$, 
because they are symmetric around the propagation axis of GWs. 
See Figure \ref{figure-curve} for the numerical plots.

The angular correlation pattern for a hemisphere has 
a rotational symmetry around the polar axis of the hemisphere. 
Therefore, the  plot for $\alpha = \pi/2$ in Figure \ref{figure-contour} 
is accidentally the same as 
the standard HD curve, 
though the former curve is for a hemisphere 
and the latter one is for the whole sky.  
The reason for this coincidence is 
that only the $\alpha = \pi/2$ has pulsars 
in the north and south skies 
in the same proportion (a half) to the whole sky in the HD curve. 

At $\gamma = 0$, the correlation decreases as $\alpha$ increases. 
This is because 
as the hemisphere is more inclined with respect to the GW axis, 
$\Theta_a$ of pulsars in the hemisphere are larger and 
hence the response $R_a$ is smaller. 

The maximum correlation $\Gamma_{\rm{max}}$ for each value of $\alpha$ 
monotonically decreases as $\alpha$ increases. 
On the other hand, 
the minimum correlation $\Gamma_{\rm{min}}$ 
monotonically increases. 
Therefore, the difference between the maximum and minimum correlations 
($\Delta\Gamma \equiv \Gamma_{\rm{max}} - \Gamma_{\rm{min}}$)  
is more sensitive to $\alpha$ than 
either of $\Gamma_{\rm{max}}$ and $\Gamma_{\rm{min}}$.

\subsection{Source localization in the sky}
\begin{figure}
\includegraphics[width=8.0cm]{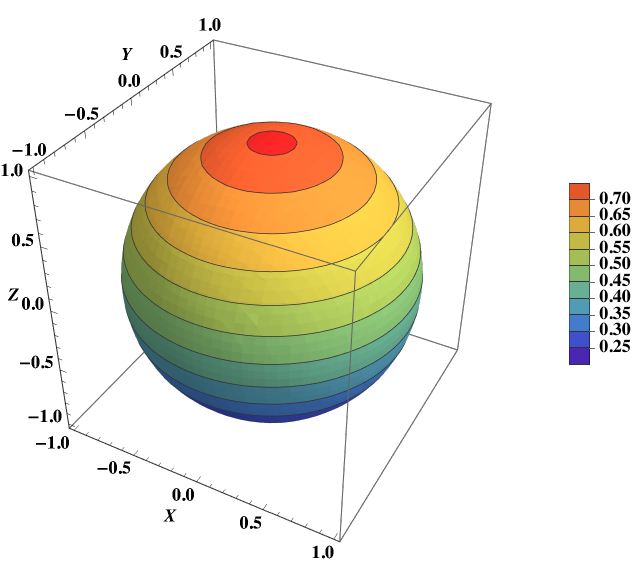}
\caption{
Contour map of the difference between the maximum and minimum correlations 
$\Delta\Gamma$. 
For its simplicity, the North Pole is chosen as the direction of a GW source. 
}
\label{figure-contour}
\end{figure}

$\Delta\Gamma$ is maximized in the
direction of the GW source on the sky. 
See Figure \ref{figure-contour} for a contour sky map 
of $\Delta\Gamma$. 
This figure shows that 
the position of  the maximum $\Delta\Gamma$ 
perfectly agrees with the GW sky location. 
The contours have no dependence on the angle $\beta$ 
as a consequence of the averaging over each hemisphere 
as mentioned above. 
Plots such as Figures \ref{figure-curve} and \ref{figure-contour} 
from future observation data could potentially 
suggest the existence of a GW compact source with an allowed region in the sky.

What is the expected accuracy of the GW sky localization 
by using the hemisphere cross-correlation? 
For its simplicity, 
we assume the measurement accuracy of the cross-correlation, 
say $10$ percents, 
which means roughly the error in $\Delta\Gamma \sim O(0.1)$. 
According to Figure \ref{figure-contour}, 
the accuracy for the determined latitude 
(corresponding to $\alpha$)
is nearly 10-20 degrees. 
This is not enough for pointing a host galaxy that may include 
a relevant GW source like a binary of supermassive black holes.

In the above discussion, we know the GW direction a priori. 
How can we use the hemisphere-averaged correlation 
for a GW source localization from future PTA observations?  
For instance, we prepare a set of hemispheres, 
e.g. 12 latitudes
and 24 longitudes (by 15 degree step). 
For $12 \times 24 = 288$ hemispheres, 
we estimate $\Delta\Gamma$.  
The maximum $\Delta\Gamma$ could 
indicate the GW location in the sky.

\subsection{On the hemisphere}
Is there a caveat in a half size of the full sky?  
In their pioneering work 
\cite{Hellings}, 
it was pointed out that 
a total number of pulsar pairs should be increased 
for statistically increasing the signal-to-noise ratio, 
because fluctuations of timings of individual pulsars are not so negligible. 
The maximum number of the pairs  
is available for the whole sky ($4\pi$).  
For $N_{\rm{pulsar}}$ as the total number of the observed pulsars in PTAs, 
the number of pulsars in the present method 
is nearly a half of $N_{\rm{pulsar}}$, 
where 
the isotropic distribution of pulsars is simply assumed. 
From the full sky to the hemisphere, 
the number of the pairs is reduced 
from $(N_{\rm{pulsar}})(N_{\rm{pulsar}}-1)/2$
to $(N_{\rm{pulsar}}/2)(N_{\rm{pulsar}}/2-1)/2$. 
For $N_{\rm{pulsar}} \gg 1$, 
this means from $(N_{\rm{pulsar}})^2/2$ to $(N_{\rm{pulsar}})^2/8$. 
Namely, it becomes one fourth, 
so that the statistical error increases twofold. 
For the current level of statistics in PTAs 
\cite{Agazie2023, Antoniadis2023, Reardon2023, Xu2023}, 
the hemisphere method may thus be a little challenging.

\subsection{Relevant GW amplitude and source distance}
How large is the GW amplitude 
relevant to the hemisphere-method? 
For estimating it, 
we use Eq. (\ref{redshift}) to obtain 
the correlation of pulsar pairs as \cite{Hellings}
\begin{align}
\langle z_a z_b \rangle 
\sim \Gamma_{\rm{H}} 
\langle h^2 \rangle ,
\label{correlation}
\end{align}
where $z_a$ and $z_b$ denote the redshifts of pulse signals 
from the a-th and b-th pulsars, respectively, 
$\langle \quad \rangle$ denotes the autocorrelation, 
and 
$\langle n_a h \rangle = 0$ and $\langle n_a n_b \rangle$. 
Roughly estimating,  
$z_a \sim \Delta t_a/T_{\rm{obs}}$, 
where $\Delta t_a$ is the delay from the expected arrival time 
and $T_{\rm{obs}}$ denotes the observation duration. 
In PTA observations, $T_{\rm{obs}}$ is roughly the inverse of 
the relevant GW frequency $f_{\rm{GW}}$.  
Hence, 
$z_a \sim z_b \sim (\Delta t_a/T_{\rm{obs}}) 
\sim f_{\rm{GW}} \Delta t_a$, 
which leads to 
the variance of  the cross correlation 
$\langle z_a z_b \rangle$ 
as 
$[\sigma(\langle z_a z_b \rangle)]^2  
\sim 
(N_{\rm pair})^{-1}
(\Delta t_a/T_{\rm{obs}})^4
$ 
for $N_{\rm pair}$ pairs of the pulsars. 
The standard deviation is roughly 
$\sigma(\langle z_a z_b \rangle) 
\sim 
(N_{\rm pulsar})^{-1}
(\Delta t_a/T_{\rm{obs}})^2$, 
where 
$N_{\rm pair} \sim (N_{\rm pulsar})^2$. 
The $\alpha$-dependence of the hemisphere correlation 
can be detected 
when 
$\sigma(\langle z_a z_b \rangle) 
\lesssim
(\delta_{\alpha} \Gamma_{\rm{H}})  
\langle h^2 \rangle$, 
where $\delta_{\alpha} \Gamma_{\rm{H}}$ 
denotes the size of the variation of $\Gamma_{\rm{H}}$ for changing $\alpha$.

Therefore, 
the GW amplitude possibly relevant to  
the current PTAs by using the hemisphere-averaged correlation  
is roughly estimated as 
\begin{align}
h &\sim \sqrt{\langle h^2 \rangle}
\notag\\
&\gtrsim 
10^{-14} 
\left(\frac{10^2}{N_{\rm{pulsar}}}\right)^{1/2}
\left(\frac{f_{\rm{GW}}}{1 \rm{yr}^{-1}}\right)
\left(\frac{\Delta t_a}{1 \mu\rm{sec.}}\right)
\left(\frac{0.1}
{\delta_{\alpha} \Gamma_{\rm{H}} }
\right)^{1/2} , 
\label{bound}
\end{align}
where 
$\delta_{\alpha} \Gamma_{\rm{H}}$ 
is $\sim 0.1$, 
and 
the average accuracy in the time residual measurement 
is $O(1) \;\mu\rm{sec}$. 

From Eq. (\ref{bound}), 
a nearby GW source 
can be marginally detected 
when the distance $D$ to the source is 
\begin{align}
D \lesssim 
&10 {\rm Mpc} 
\left(\frac{N_{\rm{pulsar}}}{10^2}\right)^{1/2}
\left(\frac{1 \mu\rm{sec.}}{\Delta t_a}\right)
\left(\frac{\delta_{\alpha} \Gamma_{\rm{H}} }
{0.1} \right)^{1/2}
\notag\\
&~~~~~
\times
\left( \frac{M}{10^9 M_{\odot}} \right)^{5/3}
\left( \frac{1{\rm yr}^{-1}}{f_{\rm GW}} \right)^{1/3} ,
\label{D}
\end{align}
where the quadrupole formula for a binary is used 
as $h \sim (MR^2)/(DT^2) \sim M^{5/3} (f_{\rm GW})^{2/3}/D$ 
for  the total mass $M$, the orbital radius $R$, the orbital period $T$ 
($T \sim 2/f_{\rm GW}$) 
\cite{CreightonBook, MaggioreBook}. 
On the other hand, 
a lower bound on the distance to the dominant source can be placed,  
unless $\delta_{\alpha} \Gamma_{\rm{H}}$ is detected. 

The near future SKA would significantly improve the sensitivity of PTAs 
by a factor $\sim 5$ or more through newly finding hundreds of millisecond pulsars 
\cite{SKA}. 
In the SKA era, the hemisphere method can be thus expected 
for e.g. the Coma cluster at $\sim 100$ Mpc.

\subsection{Multiple sky locations of GWs}
Before closing this section, 
we briefly mention the number of GW sources 
relevant to the cross-correlation curve. 
The present paper assumes a single GW source. 
How can the hemisphere cross-correlation curve be modified, 
if multiple GW compact sources are 
dominant in the cross-correlation curves?
In the forthcoming PTA observations, 
GW frequencies from multiple strong GW sources 
(if they exist) 
are likely to be distinguishable from each other, 
since it is less feasible that the orbital period of one of SMBHSs 
is accidentally the same as that of another one. 
Here, what we mean by a strong GW source is that 
it can make signals larger than the stochastic GW background. 

For each source with a different GW frequency, 
the hemisphere cross-correlation method 
can be used separately, 
if the number of the sources is not large, say a few. 
For more than a dozen of strong GW sources, 
two or more GW sources are degenerate in Fourier spaces. 
For such a case, 
the hemisphere method needs significant improvement 
by simultaneous inclusion of multiple sources. 
Detailed investigations 
taking account of two or more GW compact sources 
are beyond the scope of this paper.

Finally, we mention the term {\it within the hemisphere}.  
In the above formulation, 
most of pulsar pairs in the same hemisphere are taken into account. 
Exactly speaking, however, 
both components of a pair of pulsars
do not need live 
in the hemisphere simultaneously;  
there can be a case that one of the pair is in H but the other is outside. 
The latter case is very rare for small $\gamma$, 
while it is a larger fraction for large $\gamma$. 
On the other hand, Eq. (\ref{HD-H1}) cannot be calculated,  
unless a pulsar pair within a hemisphere is clearly defined 
in terms of the GW-oriented coordinates $(x, y, z)$.

The present result has an interesting implication. 
The distribution of pulsars observed by the current PTAs 
\cite{Agazie2023, Antoniadis2023, Reardon2023, Xu2023} 
is highly anisotropic, 
because we do not live in the galactic center. 
The deviation from the ideal HD curve in the current PTA observations 
\cite{Agazie2023, Antoniadis2023, Reardon2023, Xu2023} 
could be due to the anisotropic distribution of pulsars 
for a possible GW point-like source, 
though the deviation has not been established statistically. 
It would be interesting to pursue this direction further.


\section{Conclusion}
We discussed the hemisphere-averaged angular correlation pattern to pulsar pairs. 
Our numerical calculations showed that, 
if a  single GW source is dominant, 
the variation in a hemisphere-averaged angular correlation curve is greatest  
when the hemisphere has its North Pole at the sky location of the GW source. 
Possible GW amplitude and source distance relevant to 
the current PTAs by using the hemisphere-averaged correlation  
were investigated. 
The near future SKA will make it possible to perform sky localization 
by using  the hemisphere-averaged correlations.

We mention also the validity of the present method. 
We confirmed numerically that, 
regardless of $\alpha$ and $\beta$, 
Eq. (\ref{HD-H3}) for the whole sky perfectly recovers 
the standard HD curve, 
if $\theta$ runs from $0$ to $\pi$ in the integration by $\theta$.  

The hemisphere condition adopted in the present paper is 
that at least one pulsar of 
a pulsar pair is in the hemisphere. 
Another option is that both pulsars in a pair 
are located within the same hemisphere. 
It would be interesting to use such an option 
to examine which condition is more suitable 
for the sky localization. 
It is left for future. 

\section{Acknowledgments}
We are grateful to 
Daisuke Yamauchi, Yousuke Itoh,  
Keitaro Takahashi and Yuuiti Sendouda 
for useful discussions. 
We thank Yuya Nakamura and Ryuya Kudo 
for helpful suggestions on the numerical plots. 
This work was supported 
in part by Japan Society for the Promotion of Science (JSPS) 
Grant-in-Aid for Scientific Research, 
No. 20K03963 (H.A.).

\end{document}